\documentstyle[aps,twocolumn,epsfig,eqsecnum,rotating]{revtex}

\begin{document}

\newcommand{\mubf}{\mbox{\boldmath $\mu$}}

\twocolumn[\hsize\textwidth\columnwidth\hsize\csname    
@twocolumnfalse\endcsname                               

\begin{title} {\Large  \bf
Transition to Long Range
Magnetic Order in the Highly Frustrated Insulating Pyrochlore
Antiferromagnet  Gd$_2$Ti$_2$O$_7$}

\end{title} 
\author{N.P. Raju$^1$,  M. Dion$^{2,*}$, M. J.P.
Gingras$^{3,\dagger}$,  T.E. Mason$^{4,\ddagger}$, and J.E.
Greedan$^1$}

\address{{$^1$}Brockhouse Institute for Materials Research and
Department of  Chemistry, \\ McMaster University, Hamilton, Ontario,  L8S
4M1,  Canada}

\address{{$^2$}Department of Physics, University of Waterloo, Waterloo, 
Ontario, N2L-3G1, Canada}

\address{{$^3$}Laboratoire Louis N\'eel, Centre National de la Recherche 
Scientifique,
B.P. 166, 38042, Grenoble, Cedex, France}

\vspace{-5truemm} 
\address{{$^4$}Department of  Physics, University of
Toronto, Toronto,  Ontario, M5S 1A7, Canada}

\maketitle 

\begin{abstract}{Experimental evidence from measurements of the a.c. and 
d.c. susceptibility, and heat capacity data show that the pyrochlore structure 
oxide, Gd$_2$Ti$_2$O$_7$, exhibits short range order that starts 
developing at 30K, as well as  long range magnetic order at $T\sim 1$K. 
The Curie-Weiss temperature, $\theta_{\rm CW}$ = -9.6K,  is largely due 
to exchange interactions. Deviations from the Curie-Weiss law occur below 
$\sim$10K while magnetic heat capacity contributions are found at 
temperatures above 20K. A sharp maximum in the heat capacity at 
$T_c=0.97$K signals a transition to a long range ordered state, with the 
magnetic specific accounting for only $\sim$50\% of the magnetic entropy. 
The heat capacity above the phase transition can be modeled by assuming 
that a distribution of random fields acts on the $^8$S$_{7/2}$ ground state 
for Gd$^{3+}$. There is no frequency dependence to the a.c. susceptibility 
in either the short range or long range ordered regimes, hence suggesting
the absence of any spin-glassy behavior. Mean field theoretical calculations 
show that {\it no} long range ordered ground state exists for the conditions 
of nearest-neighbor antiferromagnetic exchange {\it and} long range dipolar 
couplings. At the mean-field level, long range order at various 
commensurate or incommensurate wave vectors is found only upon 
inclusion of exchange interactions beyond nearest-neighbor exchange and 
dipolar coupling. The properties of Gd$_2$Ti$_2$O$_7$ are compared 
with other geometrically frustrated antiferromagnets such as the 
Gd$_3$Ga$_5$O$_{12}$ gadolinium gallium garnet, 
RE$_2$Ti$_2$O$_7$ pyrochlores where RE = Tb, Ho and Tm, and 
Heisenberg-type pyrochlore such  as Y$_2$Mo$_2$O$_7$, 
Tb$_2$Mo$_2$O$_7$, and spinels such as ZnFe$_2$O$_4$}	

\vspace{1cm}


\end{abstract}

\vskip2pc]                                              

 
\section{Introduction}
\label{Intro}

\indent There has been in the past eight years an enormous amount of
theoretical and experimental activity devoted to the study of
highly geometrically frustrated antiferromagnetic 
materials~\cite{review}.  The main reason for
this interest stems from the suggestion that a high degree of frustration
can induce sufficiently large zero-temperature quantum spin
fluctuations as to destroy long range N\'eel order even in three
dimensions. This could give rise to new exotic and intrinsically quantum
mechanical magnetic ground states such as dimerized ground states,
``spin nematics'', or  fully disordered states (e.g. RVB-like) with no
broken spin or lattice symmeties~\cite{chandra,huillier,lacroix}.  
Frustration
arises when a magnetic system cannot minimize its total classical
ground-state energy by minimizing the bond energy of each spin-spin
interaction individually~\cite{toulouse}. This is the case, for
example, in systems where antiferromagnetically coupled spins reside on
a network made of basic units such as triangles or tetrahedra.
On a triangular plaquette, vector (i.e. XY or Heisenberg)
spins can manage the frustration better than Ising moments by adopting
a noncollinear structure with the spins making an angle of 120$^\circ$
from each other.   Triangular or tetrahedral units
can be put together to form a regular lattice
such that they are either {\it edge sharing} or
{\it corner sharing}. For example, the space-filling arrangements of
edge-sharing triangles and tetrahedra form the well-known triangular
and face-centered cubic lattices in two and three dimensions,
respectively.  In two dimensions, a network of corner-sharing triangles
forms the kagom\'e lattice~\cite{review,chandra,huillier}.
In three dimensions, a lattice of
corner-sharing tetrahedra forms the structures found in spinels, Lave
phases, and pyrochlore crystals~\cite{lacroix,coey,villain}, while
corner-sharing triangles give the familiar 
garnets~\cite{ggg_old,ramirez_ggg1,ramirez_ggg2,petrenko,schiffer}. 

Among highly frustrated antiferromagnets, the
three-dimensional pyrochlore lattice of corner-sharing 
tetrahedra is a particularly interesting system (see Fig. 1).
Theory~\cite{villain,reimers-mft,moessner} and Monte Carlo
simulations~\cite{moessner,reimers-pyro} show that
classical Heisenberg spins residing on the vertices of the pyrochlore
lattice and interacting only via nearest-neighbor antiferromagnetic
exchange do not show a transition to long range magnetic order at
nonzero temperature.  This is very different from the, also frustrated,
classical nearest-neighbor fcc Heisenberg antiferromagnet where long range 
order
occurs at finite-temperature via a 1st order transition driven by
thermally-induced order-by-disorder~\cite{FCC,larson,heinila}. 
Because of their failure to order even at the classical level,
the high-frustration present in
pyrochlore antiferromagnets would appear to make these systems
excellent candidates to search for novel
three-dimensional quantum disordered  magnetic ground states.  Indeed, 
recent numerical calculations suggest that the quantum $S=1/2$
pyrochlore Heisenberg antiferromagnet may be a quantum 
spin-liquid~\cite{lacroix}.

In all real systems there exist perturbations, $\{H'\}$, beyond the
nearest-neighbor Heisenberg Hamiltonian such as further than
nearest-neighbor exchange, single-ion and exchange anisotropy, and
magnetic dipolar couplings. For a classical system, one generally expects 
that
such perturbations will select a unique classical ground state
to which a transition at nonzero temperature can occur. It is also
possible that
the energetic perturbations $\{H'\}$ can sufficiently reduce the
classical degeneracy such that ``additional'' order-by-disorder via
thermal~\cite{villain,reimers_plumer-FeF3,bramwell-obdo} 
and/or weak
quantum fluctuations~\cite{sachdev}, such as occur in the fcc
antiferromagnet~\cite{FCC,larson,heinila}, can ``complete''
the ground-state selection and give rise to a transition to conventional
long range N\'eel order at nonzero
temperature.  However, in presence of quantum fluctuations (i.e. finite
$S$ spin value), one expects that for small spin value $S$
and/or sufficiently weak $\{H'\}$, a quantum disordered
phase may occur~\cite{lacroix,sachdev}.

What is perhaps one of the most interesting issues in geometrically 
frustrated 
antiferromagnet systems, is that a large number of
highly frustrated antiferromagnetic insulators exhibit
spin-glass behavior rather spin-liquid behavior. For example, in a number of
antiferromagnetic pyrochlore oxydes, such as 
Y$_2$Mo$_2$O$_7$~\cite{dunsiger,gingras_ymoo,gardner_y2mo2o7} 
and
Tb$_2$Mo$_2$O$_7$~\cite{dunsiger,gaulin}, spin-glass
behavior is observed similar to what is seen in conventional randomly
disordered and frustrated spin glass materials~\cite{byfh}, even though 
the measured disorder level is immeasurably small in the 
pyrochlores~\cite{greedan_disorder1,greedan_disorder2}. 

Pyrochlores oxides, of which two examples have just been cited, present
a number of opportunities for studying
 geometrically frustrated antiferromagnets.
In this structure both cation sites
in Fd$\bar 3$m, the 16c site normally occuped by a transition element and 
16d normally
occupied by a lanthanide, have the same ``pyrochlore'' topology, i.e.,
the three dimensional network of corner sharing tetrahedra shown in Figure 
1. Thus materials exist with only the 16c site magnetic 
(Y$_2$Mo$_2$O$_7$)~\cite{dunsiger,gingras_ymoo,gardner_y2mo2o7}, 
with both
sites magnetic (Tb$_2$Mo$_2$O$_7$)~\cite{dunsiger,gaulin},
and with only the 16d site magnetic, the 
RE$_2$Ti$_2$O$_7$ materials for example. This latter series has attracted
attention very recently with studies for RE= Tm, Ho and 		
Tb~\cite{tm2ti2o7_1,tm2ti2o7_2,ho2ti2o7,spin-ice,orbach,tb2ti2o7,tb2ti2o7_prop}.
The RE$_2$Ti$_2$O$_7$ materials are generally quite well-ordered
crystallographically, with oxygen non-stoichiometry and 16c/16d site
cation admixing at or below the limit of detection by neutron 
diffraction~\cite{greedan_disorder2}.

Gd$_2$Ti$_2$O$_7$ has not yet been studied in detail and there exist 
some
compelling reasons to do so~\cite{gd2ti2o7_old}. 
Previous reports for Gd$_2$Ti$_2$O$_7$ indicate no
long range order down to 1K~\cite{gd2ti2o7_old}.
The Gd$^{3+}$ ion ($4f^7$) is spin-only with 
a $^8S_{7/2}$ ground state and thus, crystal field splittings and anisotropy,
which play a large role in the properties of the aforementioned Tm$^{3+}$,
Ho$^{3+}$ and Tb$^{3+}$ materials, as will be discussed later, are 
expected to be relatively
unimportant. Gd$_2$Ti$_2$O$_7$ then, should be an excellent 
approximation to 
a classical Heisenberg antiferromagnetic system with dipole-dipole
interactions as leading perturbations $\{H'\}$. In addition it is important to 
compare this material to the Gd$_3$Ga$_5$O$_{12}$, gadolinium gallium 
garnet (GGG),
where the Gd$^{3+}$ ions reside on a three-dimensional sublattice of 
corner-sharing
triangles~\cite{ggg_old,ramirez_ggg1,ramirez_ggg2,petrenko,schiffer}. 
GGG has been found to possess a very unusual set of 
thermodynamic properties with anomalous specific heat behavior, spin
glass magnetic properties and no true long range 
order~\cite{ramirez_ggg1,ramirez_ggg2},
but incommensurate
short range order developing at very low
temperatures~\cite{petrenko}.   Interesting properties are also observed for 
applied magnetic fields in the range $[0.1-0.7]$ Tesla~\cite{schiffer}.
same time Monte Carlo simulations for GGG have found some intrinsic
(e.g. disorder-free) glassy behavior with no evidence for the development
of short range order~\cite{ramirez_ggg1,petrenko}.
Hence, one further motivation for studying  Gd$_2$Ti$_2$O$_7$ was to 
hope bridge a gap
between the peculiar behavior of GGG and the transition metal pyrochlores, 
such
as Y$_2$Mo$_2$O$_7$ and Tb$_2$Mo$_2$O$_7$ which show
spin-glass behavior and, hence,  possibly indirectly
gain some insight on the thermodynamic behavior of both GGG and the
insulating antiferromagnetic Heisenberg pyrochlore systems.

In this work a detailed study of Gd$_2$Ti$_2$O$_7$ has been carried out
including both a.c. and d.c. susceptibility and heat capacity studies.
To complement the experimental work, results from mean-field
theoretical calculations are presented which take into account exchange
and dipolar interactions.


\section{Experimental Method}
\label{exp_method}

\subsection{Sample Preparation}

A polycrystalline sample of Gd$_2$Ti$_2$O$_7$ was prepared by high 
temperature
solid state reaction.  Starting materials, Gd$_2$O$_3$ and TiO$_2$,
were taken in stoichiometric proportions and mixed thoroughly.  The
mixture was pressed into pellets and heated in an alumina crucible at
1400C in air for 12 hours.  The powder x-ray diffraction pattern of the
sample obtained using a Guinier-Hagg camera indicate that the sample
formed is single phase with the cubic pyrochlore structure. The size of the 
conventional cubic unit cell is $a_0=10.184(1)\AA$.

\subsection{DC \& AC Magnetic Susceptibility Measurements}

The DC magnetic susceptibility, $\chi$, was measured using a SQUID
magnetometer (Quantum Design, San Diego) in the temperature range
2$-$300K.  The AC susceptibility, $\chi_{\rm AC}$, was measured at
different frequencies by the mutual inductance method.  The primary
coil of the mutual inductor is energized by a frequency generator (DS
335, Stanford Research Systems) and the output across the two identical
secondary coils, wound in opposite directions, was measured using a
lock-in-amplifier (SR-830 DSP, Stanford Research Systems).  The sample
susceptibility was determined from the difference in the outputs with
the sample in the middle of the top secondary coil and without the
sample.  The cryostat used for the temperature variation is described
in the section below.

\subsection{Specific Heat Measurements}

The specific heat of the sample in the form of a pellet ($\approx$100
mg) was measured in the temperature range 0.6$-$35K using a
quasiadiabatic calorimeter and a commercial Heliox sorption pumped
$^3$He cryostat supplied by Oxford Instruments.  The sample was mounted
on a thin sapphire plate with apiezon for better thermal contact.
Underneath the sapphire plate a strain gauge heater and a RuO$_2$
temperature sensor were attached with G-E varnish.  The temperature of
the calorimetric cell was controlled from the $^3$He pot on the
Heliox.  The sample temperature was measured using an LR-700 AC
resistance bridge at a frequency of 16Hz.  The specific heat of the
sample was obtained by subtracting the contribution of the addendum,
measured separately, from the total
measured heat capacity.

\section{Experimental Results} \label{exp_results}

The DC susceptibility, Fig. 2a, $\chi$, measured at an applied field of 0.01
Tesla vs temperature is found to obey the Curie-Weiss behavior in the
range 10K$-$300K. An
effective magnetic moment of 7.7$\mu_{\rm B}/{\rm Gd}^{3+}$
obtained from the Curie-Weiss
fit is close to the expected value of 7.94$\mu_{\rm B}/{\rm Gd}^{3+}$
for the free ion,
$^8$S$_{7/2}$, and a paramagnetic Curie temperature, $\theta_{\rm 
CW}$,
of -9.6(3)K indicates antiferromagnetic interactions between the
Gd$^{3+}$ spins.
It is worth noting that $\chi$ starts deviating at a temperature of
the order of $\theta_{\rm CW}$ as it ought to be for a ``conventional''
system undergoing a transition to long range order.
That $\theta_{\rm CW}$
is predominantly
due to exchange interactions as opposed to 
crystal field effects is confirmed by measurements
on the magnetically diluted system
(Gd$_{0.02}$Y$_{0.98}$)$_2$Ti$_2$O$_7$, for which $\theta_{\rm
CW}$ is much reduced and of the order of $\sim -0.9$K (Fig. 2b).
The absence of any magnetic ordering down to 2K in the
concentrated system, even though $\theta_{\rm CW}$ is about five times
larger than this temperature, suggests the presence of important magnetic
frustration inhibiting the occurrence of magnetic long range
order.	

In search of a possible magnetic ordering below 2K, AC susceptibility,
$\chi_{\rm AC}$, was measured down to 0.3K.  The temperature variation
of $\chi_{\rm AC}$ for different frequencies, Fig.  3, exhibits two
features, a broad peak centered at about 2K and a sharp down turn below
about 1K, the latter possibly signaling a transition to long range
antiferromagnetic order.
$\chi_{\rm AC}(\omega)$ appears to be independent of frequency which
would seem to rule out a spin glass state, as opposed to what has been
found in other pyrochlore oxydes such as
Y$_2$Mo$_2$O$_7$~\cite{dunsiger,gingras_ymoo,gardner_y2mo2o7},
Tb$_2$Mo$_2$O$_7$~\cite{dunsiger,gaulin} and the frustrated
Gd$_3$Ga$_5$O$_{12}$ 
garnet~\cite{ramirez_ggg1,ramirez_ggg2,schiffer}.

The specific heat, $C_p$, as a function of temperature is shown in Fig.
4.  There is a broad peak centered around 2K and a very sharp peak
slightly below 1K indicating the presence of short range correlations
and, in agreement with the AC susceptibility data, the development of
long range magnetic order via a sharp transition at 1K.  The solid
line corresponds to the estimated lattice specific heat, $C_l$, of
Gd$_2$Ti$_2$O$_7$ determined by scaling the specific heat for
Y$_2$Ti$_2$O$_7$, which is insulating, non-magnetic, and isostructural
to Gd$_2$Ti$_2$O$_7$.  The magnetic specific heat, $C_m$, was obtained
by subtracting $C_l$ from $C_p$ and its temperature variation is shown
in Fig.5.

The Gd$^{3+}$ ion has an isotropic spin  of $S=7/2$ with no orbital
magnetic moment contribution and the degeneracy of the 8$-$level ground
state cannot be lifted by the crystal electric field beyond a fraction
of a Kelvin.  The presence of $C_m$ up to about 30K clearly indicates
that the ground state degeneracy is lifted by magnetic interactions.

The magnetic entropy, $S_m$, was obtained by extrapolating the $C_m/T$
behavior to 0K and numerically integrating it vs temperature.  The
total magnetic entropy is 33.8 J mol-1K-1 which is close to the
expected 2Rln(8) $=$34.6 Jmol$^{-1}$K$^{-1}$ for an $S=7/2$ system.
The entropy recovered at the long range order temperature is about 50\%
of the total value which indicates that a sizeable fraction 
of the entropy is
due to the short range correlations present above $T=1$K.

An attempt was made to fit the $C_m$ data above 1K.  A zeroth order
model consisting of a simple Schottky anomaly based on a splitting
scheme of 8 equally spaced discrete levels, i.e., assuming a unique
value for the internal magnetic field at each Gd$^{3+}$ site, gives a poor
fit, as might be expected.  A much better fit is obtained by
assuming a continuous range of energy level splittings with a truncated
Gaussian distribution. The probability distribution is normalized such
that the area under the curve is unity.  The resulting fit is shown in
Fig. 5.  This
model is equivalent to assuming a distribution of internal magnetic
fields, i.e., an array of random fields as appropriate to a thermal
regime dominated by quasi-static 
short range magnetic order.  A similar approach has
been used before to model the specific heat anomaly due to the Gd 
sublattice
in Gd$_2$Mo$_2$O$_7$~\cite{gd2mo2o7-sg}.  In this pyrochlore 
structure
material the Mo$^{4+}$ sublattice undergoes a spin glass type of order
at 60K while the Gd$^{3+}$ specific heat contribution is also a broad
peak centered about 9K.

In summary, experimental results obtained  from AC
and DC susceptibility measurements as well as specific heat
measurements reported in Section~\ref{exp_results} give strong compelling
evidence for a single sharp transition to a long range ordered state at
$T_c=0.97$K preceeded by a short range ordered regime which extends to 
approximately
30$T_c$ $\sim 3\theta{\rm CW}$. The heat capacity in this regime can be 
modelled in terms of a 
distribution of random exchange fields acting on the $^8$S$_{7/2}$ ground 
state
of Gd$^{3+}$.

\section{Mean-Field Theory}
\label{theory}

\subsection{Model and Method}

Our aim in this section is to determine, within mean-field theory,
what the expected magnetic properties and type of magnetic ordered 
phase(s)
for a classical spin model of Gd$_2$Ti$_2$O$_7$ are.

We first consider the following classical spin Hamiltonian for 
Gd$_2$Ti$_2$O$_7$:
\begin{equation}
H=\frac{1}{2}\sum_{(i,j)}-J_{ij}{\bf S_i}\cdot {\bf S_j} \;\; +
 \frac{1}{2}\sum_{(i,j)}
\left ( \frac{\mubf_i\cdot{\mubf}_j}{r_{ij}^3}
-3\frac{\mbox{\boldmath $\mu$}_i\cdot{\bf  r}_{ij}{\bf r}_{ij}\cdot
{\mubf}_j}{r_{ij}^5} \right)
		\;\;\;. 
\end{equation}
The first term is the isotropic Heisenberg  exchange interaction, and the
second term is the dipolar coupling between the Gd magnetic moments.
For the open pyrochlore lattice structure, we expect very small 
second and further nearest-neighbor exchange coupling,  $J_{n\ge 2}$,
compared to the nearest-neighbor $J_1$ ($J_{n\ge 2}< 
0.05J_1$)~\cite{greedan_open_pyro}.
Hence, we first consider the case where the sum in the first
(exchange) term of Eq.(4.1) 
above is restricted to the nearest-neighbor exchange $J_1$ 
only~\cite{dion_gd2ti2o7}.

Gd$^{3+}$ has a spin $S=7/2$, which gives an effective
 dipole moment of $\mu({\rm Gd}^{3+})=
g\mu_{\rm B}{\sqrt{S(S+1)}}= 7.94\mu_{\rm B}$, with $g=2$, in
good agreement with the Curie constant determined in Section III. This 
gives an estimate
for the nearest-neighbor strength of the dipole-dipole interaction
$D_{dd}=63\mu_{\rm B}^2 \mu_0/(4\pi r_{nn}^3)$, where $\mu_0$ is the
magnetic permeability.
With a Gd$^{3+}$ at ${\bf r}=(0,0,0)$ and a nearest-neighbor at
${\bf r}_{nn}=(a/4,a/4,0)$, where $a=10.184\AA$ is the size of the 
conventional
cubic unit cell, we find $D_{dd}\approx 0.84$K.
 
An estimate of the nearest-neighbor exchange $J_1$ can be found from the
measured Curie-Weiss temperature (see below). We have $\theta_{\rm 
CW} \sim
-9.6$K.  This gives for the effective classical nearest-neighbor
exchange, $J^{\rm cl}_1 = J_1 S(S+1) \sim -4.8$K using 
$\theta_{\rm CW}=zJ_1S(S+1)/3$, where $z=6$ is the number of nearest-
neighbor.
We henceforth use a classical approximation of Eq. (2.1) above, where we
use unit length 
vectors ${\bf S}_i$, and replace 
${\bf \mu}_i$ by ${\bf S}_i(D_{dd})^{1/2}$, and express $r_{ij}$ in
units of the nearest-neighbor distance. We
set $J_1=-4.8$K and a strength of $0.84$K for the nearest-neighbor dipole
coupling. Below, we take $D_{dd}/J_1=0.2$~\cite{dipole_theta_cw}.
Hence, unlike in the transition metal pyrochlores,
dipole-dipole interactions is the major perturbation at play
beyond the nearest-neighbor Heisenberg exchange coupling
in Gd$_2$Ti$_2$O$_7$.

We now proceed along the lines of Reimers, Berlinskly and Shi
in their mean-field study of 
Heisenberg pyrochlore antiferromagnets~\cite{reimers-mft}.
We consider the mean-field order parameters, ${\bf  B}({\bf  r_i})$ at
site ${\bf r}_i$, in terms of Fourier components.
The pyrochlore lattice is a non-Bravais lattice, and we
use a rhombohedral basis where there are four
atoms per unit cell located at
$(0,0,0), (1/4,1/4,0), (1/4, 0, 1/4)$, and
$(0,1/4,1/4)$ in units of the conventional cubic unit cell. 
We relabel the spins, ${\bf S}({\bf r}_i)$, in terms of unit cell coordinates, 
and a
sublattice index within the unit cell, and
take advantage  of the translational symmetry of the lattice, and
expand the order parameters ${\bf B}(\bf r_i)$ in terms of Fourier
components.
In this case ${\bf  B}^a({\bf  r}_i)$ on the $a$'th sublattice site of the
unit cell located at ${\bf r}_i$
can be written as
\begin{equation}
{\bf B}^a({\bf }{\bf r}_i) = \sum_{{\bf q}} 
{\bf B}^a({\bf q})\exp(i{\bf q}\cdot {\bf  }_i)	\;\;.
\end{equation}

\vspace{2cm}

The spin-spin interaction matrix, ${\cal J}_{\alpha\beta}(\vert {\bf 
r}_{ij}\vert)$,
including both exchange and dipolar interactions, reads:
\begin{equation}
{\cal J}_{\alpha\beta}^{ab}(\vert{\bf r}_{ij}\vert )= 
J_1\delta_{\alpha\beta}\delta_{r_{ij},r_{nn}} +
D_{dd}\left \{
\frac{ \delta_{\alpha\beta} } {(r_{ij}^{ab})^3} - 					
3
\frac{ r_{ij,\alpha}^{ab}r_{ij,\beta}^{ab} } {(r_{ij}^{ab})^5}
\right \} 
\;,
\end{equation}
where $\delta_{\alpha\beta}$ is the Kronecker delta, and
$\alpha$ and $\beta$ refer to the $x,y,z$ cartesian
components of ${\bf S}_i$ and
${\bf r}_{ij}^{ab}$.
$r_{ij,\alpha}^{ab}$ denotes the $\alpha$ components of the
interspin vector ${\bf r}_{ij}$ that 
connects spin ${\bf S}_i^a$ to spin ${\bf S}_j^b$.
We write ${\cal J}_{\alpha\beta}$
in terms of its
Fourier components as
\begin{equation}
{\cal J}_{\alpha\beta}^{ab}(\vert {\bf r}_{ij}\vert)=\frac{1}{n}\sum_{\bf 
q}
{\cal J}_{\alpha\beta}^{ab}({\bf q})\exp(-i{\bf q}\cdot {\bf r}_{ij})\;\;
	.
\end{equation}
where $N$ is the number of unit cells with 4 spins per unit cell.

The quadratic part of the mean-field free-energy, $F^{(2)}$,
then becomes~\cite{reimers-mft}:
\begin{equation}
F^{(2)}(T)/N=\frac{1}{2}\sum_{{\bf q},(ab),(\alpha\beta)}
\!\!\!\!\!\!\! B^a_\alpha({\bf q}) 
\left\{ 3T\delta_{ab}\delta_{\alpha\beta}-{\cal J}_
{\alpha\beta}^{ab}({\bf q})\right\}B^b_\beta(-\bf q) 
\;\;	.
\end{equation}

Diagonalizing $F^{(2)}(T)$ requires transforming to normal modes of the 
system
\begin{equation}
B_\alpha^a({\bf q}) = \sum_i\sum_\beta 
U_{\alpha\beta}^{a,i}\Phi_{\beta}^i(\bf q)
\end{equation}
where $\Phi_{\beta}^i(\bf q)$ are the eigenmodes, and
$U(\bf q)$ is the unitary matrix that diagonalizes ${\cal J}(\bf  q)$
in the spin$-$sublattice space, with eigenvalues $\lambda(\bf q)$
\begin{equation}
\sum_b\sum_\beta {\cal J}_{\alpha\beta}^{ab}({\bf 
q})U^{bi}_{\beta\gamma}({\bf  q})=
\lambda_\gamma^i({\bf q})U_{\alpha\gamma}^{ai}({\bf  q})
\;\; .
\end{equation}
Henceforth we will use
the convention that indices $(ab)$ label sublattices,
that indices $(ijk)$ label the normal modes, and that $(\alpha\beta\gamma)$
label spin components.
We express $F^{(2)}(T)$ in terms of normal modes
\begin{equation}
F^{(2)}/N=\frac{1}{2}\sum_{\bf  q} \sum_i \sum_\gamma
\Phi_{\gamma}^i({\bf  q})\Phi_\gamma ^i(-{\bf  q})
\left\{ 3T-\lambda_\gamma^i({\bf q}) \right \}
\end{equation} 
The first ordered state of the system occurs at the temperature 
\begin{equation}
T_c = \frac{1}{3}\max_{{\bf q},i,\alpha} \{\lambda_\alpha^i({\bf q})\}
\end{equation}
where max$_{{\bf q},i,\alpha} \{\lambda_\alpha^i({\bf q})\}$ indicates a 
global maximum of the spectrum of $\lambda_\alpha^i({\bf q})$ for all 
${\bf q}$.

Let us briefly explain how we procede using the above set of
equations to determine the
``soft mode(s)'' of the system at $T_c$. 
Firstly, the Fourier transform of ${\cal}J^{ab}_{\alpha\beta}(\vert
{\bf r}_{ij}\vert)$ is calculated using Eq. (4.3)~\cite{range_dipoles}.
For the rhombohedral basis used
above, the space is of dimension ${\cal D}_{\bf S}\otimes {\cal D}_{\rm 
sl}$,
where the spin-component subspace, ${\cal D}_{\bf S}$, is of dimension
${3\!\times\!3}$ and the sublattice subspace, $ {\cal D}_{\rm sl}$, is of
dimension ${4\!\times\! 4}$. The eigenvalues, $\{\lambda_\alpha^i(\bf 
q)\}$,
and eigenvectors $\Phi_{\alpha}^i(\bf q)$ are determined by reshaping
${\cal J}^{ab}_{\alpha\beta}({\bf q})$ into a ${12\!\times\!12}$ array.
The pyrochlore lattice has a symmetry of inversion with respect to a
lattice point and this implies that ${\cal J}^{ab}_{\alpha\beta}({\bf q})$
is real and symmetric.
The eigenvalues and eigenvectors are found using a standard 
numerical packages for eigen problems of real symmetric matrices.

\subsection{Results}

For $D_{dd}=0$, we recover the results of Ref.~\cite{reimers-mft}.
Before we present the results with the dipolar interactions,
we review what the mean-field results
found for the isotropic pyrochlore class Heisenberg antiferromagnet
depending on the values of the second, $J_2$ and third, $J_3$ 
nearest-neighbor exchange couplings are~\cite{reimers-mft}.
For $J_2=J_3=0$ there are two dispersionless unstable or critical modes
throughout the Brillouin zone. 
There are therefore no selected wavevector for long range order. Numerical
work has shown that no long range order occurs at nonzero temperature in
the nearest-neighbor classical Heisenberg
 pyrochlore antiferromagnet~\cite{moessner,reimers-pyro}.
For $J_3=0$, ferromagnetic 
$J_2>0$ gives rise to an ordering at an incommensurate wavevector, while
for antiferromagnetic $J_2<0$, the system orders at ${\bf q}^*=0$.
For $J_2=0$, and ferromagnetic $J_3>0$, the system also orders at 
${\bf q}^*=0$, while there are dispersionless (degeneracy lines) along
certain symmetry directions for $J_2=0$ and antiferromagnetic $J_3<0$.
In the overall parameter space $\{J_2/J_1,J_3/J_1\}$, long range
order is always expected to occur at nonzero temperature within mean-field
theory, except for $J_2\equiv 0$ and antiferromagnetic $J_3\le 0$ (Fig. 6
in Ref.~\cite{reimers-mft}).

We now consider the case where $D_{dd}/J_1=0.2$, and
first set $J_2=J_3=0$. Naively, one might have thought that
(i) the long range and (ii) anisotropic nature of the dipolar interactions
would lift {\it all} macroscopic ground state degeneracies that occur in 
the isotropic nearest-neighbor ($J_1<0$) Heisenberg antiferromagnet and 
give rise to a unique selected wavevector ${\bf q}^*$, at which long range 
order would occur~\cite{reimers-mft}.
This is not the case.  We find that the largest eigenvalue
$\lambda_\alpha^i({\bf q})$ that controls the mean-field $T_c$
(Eq. 4.9) is dispersionless along the star of the $[111]$ direction 
in the cubic basis (Fig. 6). The figure shows
$\lambda_{\rm max}({\bf q})$ as a function of ${\bf q}_1$ in the $[110]$ 
and
${\bf q}_2$ in the $[001]$, where $\lambda_{\rm max}({\bf q})$ is the 
largest eigenvalue of $\lambda_\alpha^i({\bf q})$ at a given ${\bf q}$. 
Hence, no-long range order is to be expected
in this system within the mean-field approximation~\cite{range_dipoles}.
In this context, in it interesting to note that the combined
{\it long range} dipolar {\it and} RKKY interactions
in the problem of nuclear magnetism in Cu and Ag do not lead either to
a full selection of a unique classical long range ordered state
below the mean-field $T_c$~\cite{CuAg_order}.
Such ``degeneration lines''
as found in the present system also occur in other frustrated systems such
as the nearest-neighbor Heisenberg fcc antiferromagnet where there
are degeneration lines along the $\pi/a(1,q,0)$ 
direction~\cite{FCC,larson,heinila}. Degeneration (spiral) lines
also occur in the more complicated case of the rhombohedral 
antiferromagnet~\cite{rastelli}. By analogy with the work
on the 
frustrated fcc~\cite{FCC,larson,heinila,CuAg_order} and 
rhombohedral~\cite{rastelli}
antiferromagnets, we expect that for deneration lines (as opposed to
degeneration {\it zone} as in the case of the nearest-neighbor
pyrochlore antiferromagnet~\cite{reimers-mft,moessner}),
Thermal and/or quantum fluctuations will
restore long range order at finite temperature via a process of 
order-by-disorder. Work in that direction is in progress and will
be reported elsewhere~\cite{dion_gd2ti2o7}.

We find that for either nonzero ferromagnetic or antiferromagnetic,
$\vert J_2\vert \ll D_{dd}$ and/or $\vert J_3\vert \ll D_{dd}$, 
that the line-degeneracy along the $[111]$ direction is lifted
and that a specific value $q^*$ along that direction is picked-up,
giving rise to an absolute maximum of $\lambda_\alpha^i({\bf q})$.
For sufficiently large $\vert J_2\vert$ and/or $\vert J_3\vert$ compared
to $D_{dd}$, a different selected wavevector direction is
chosen as found by Reimers et al.~\cite{reimers-mft}, except that here,
there is no degeneration line occuring for $J_2=0$ and $J_3<0$, as found
in Ref.~\cite{reimers-mft} when
$D_{dd}\ne 0$, $D_{dd}\ll \vert J_2\vert$ and
$D_{dd}\ll \vert J_3\vert$.  In other words, all non-global degeneracies are 
lifted in the case where $D_{dd}\ne 0$, $J_2\ne 0$, and $J_3\ne 0$.

In summary, we would expect that long range order should occur
in Gd$_2$Ti$_2$O$_7$, either via an order-by-disorder mechanism,
or via energetic selection of an ordering wavevector via 
superexchange couplings beyond $J_1$ and dipolar interactions,
D$_{dd}$.

%
\section{Discussion}
\label{discussion}

It is  useful to compare Gd$_2$Ti$_2$O$_7$ 
with related systems such as the
remaining RE$_2$Ti$_2$O$_7$ materials, Gd$_3$Ga$_5$O$_{12}$ 
(GGG),  Gd$_2$O$_3$, cubic-Gd$_2$O$_3$ (C--Gd$_2$O$_3$) and also 
GdAlO$_3$ for reasons which should soon become clear.
 
C-Gd$_2$O$_3$ crystallizes in the so-called
bixbyite structure, Ia3, with two distinct crystallographic
sites. However, the sublattice of the
two sites taken together is an excellent approximation to an fcc lattice. Its
properties are as follows. Curie-Weiss behavior is observed with 
$\theta_{\rm CW}$
= -17K but magnetic order of undetermined range is not found down to
1.6K~\cite{C-Gd2O3}.
In fact the description of the neutron diffraction data for
C-Gd$_2$O$_3$ bears a striking resemblance to that for 
GGG~\cite{petrenko}.
The heat capacity from 1.4K to 18K shows only a broad ``Schottky'' type 
anomaly peaked near 2K with magnetic contributions extending to about
20K~\cite{C-Gd2O3}.  Neutron diffraction data in the form of diffuse 
scattering
confirm that the short range magnetic correlations do extend to at 
least 20K~\cite{C-Gd2O3}.
It is clear that C-Gd$_2$O$_3$ should be reconsidered as a
geometrically frustrated antiferromagnet material although 
the original interpretation of its properties
was not presented in those terms. In particular it is important to determine
whether C-Gd$_2$O$_3$ does indeed undergo true long range order as 
seems to 
be the case for Gd$_2$Ti$_2$O$_7$
and if the susceptibility is frequency dependent.

GGG has been recognized as a geometrically frustrated antiferromagnet
since 1979~\cite{ggg_old}.
Its properties
are an amalgam of those of Gd$_2$Ti$_2$O$_7$
and C-Gd$_2$O$_3$ scaled to lower energies.
For example $\theta_{\rm CW}$ = -2.3K and the maximum in the heat 
capacity
occurs at 0.8K. True long range order is not established down to 43mK
but incommensurate short range order on a $\sim 100\AA$ length scale
is found at the lowest temperature studied~\cite{petrenko}.  The extended 
short range order starts developing rapidly at $\sim$ 140 mK,  which is 
close to the temperature at which the nonlinear coefficient $\chi_3(T)$ 
peaks as found in Ref.~\cite{ramirez_ggg2}.
Coexistence of spin glass and long range order 
in the form of a frequency dependent susceptibility is found,
which contrasts sharply with the lack thereof in Gd$_2$Ti$_2$O$_7$.
At this point it is
useful to point out that Gd$_2$Ti$_2$O$_7$, GGG and C-Gd$_2$O$_3$ 
represent anomalies in the 
context of magnetic Gd$^{3+}$ oxides. For example monoclinic 
Gd$_2$O$_3$ in which
the Gd sublattice is not fcc, unlike C-Gd$_2$O$_3$,  is a normal 
antiferromagnet with T$_c$= 3.8K
and greater than 90\% entropy removal below T$_c$~\cite{Gd2AlO3}.
In GdAlO$_3$ the Gd ions
are on a simple cubic lattice and $\theta_{\rm CW}$ = -4.8K, 
T$_c$ = 3.8K and nearly 100\% of the entropy is removed below 
T$_c$~\cite{Gd2AlO3}.

At present the strong contrast in behavior between Gd$_2$Ti$_2$O$_7$
and GGG remains
unexplained. On the subject of why Gd$_2$Ti$_2$O$_7$
orders and GGG does not, one can
only speculate. For example, for the very specific topology of the garnet
lattice, there may remain degenerate or quasi degenerate dispersion lines
or surfaces of zero mode in $q$-space which survive even upon the 
inclusion
of perturbations such as dipolar and or higher than first neighbor
exchange interactions. Thus the selection of an ordering wave vector,
q$^*$, for GGG may be much less robust than for the much different 
topology
presented by Gd$_2$Ti$_2$O$_7$ for perturbations 
$\{H'\}$ of similar order of magnitude~\cite{dion_gd2ti2o7}.

Returning to the RE$_2$Ti$_2$O$_7$ series, as mentioned, 
Gd$_2$Ti$_2$O$_7$
offers the
opportunity to study a system in which the crystal field and anisotropy
perturbations are minimized. As this is certainly not the case for
RE= Tb, Ho and Tm, some comment on the symmetry of the local 
environment
at the RE site is in order. The 16d rare-earth 
site is coordinated by two sets of 
oxygen atoms, six O1(48f) and two O2(8b), giving eight-fold coordination
overall.  It is important to note that the RE site symmetry is strongly 
distorted from cubic, which would imply eight equal RE-O distances (for 
RE=Gd the sum of the ionic radii give 2.42 $\AA$) and O-RE-O angles of 
70.5$^\circ$, 109.5$^\circ$ and 180$^\circ$. In RE$_2$Ti$_2$O$_7$, the 
six O1 atoms form a puckered ring about the RE (Gd-O distance of 
2.55$\AA$) and
the two O2 atoms a linear O2-RE-O2 unit oriented normal to the mean 
plane of the puckered ring with extremely short RE-O2 distances(Gd-O 
distance is 2.21$\AA$). This Gd-O distance is among the shortest, if not the 
shortest,
such distance known in Gd oxide chemistry and implies a very strong
interaction. This observation suggests that a crystal field of axial 
symmetry might be an even better approximation than cubic. The
O2-Gd-O2
angle is of course 180$^\circ$ while the O1-Gd-O1 angles are 62.7$^\circ$,
117$^\circ$  and 180$^\circ$.  and the O1-Gd-O2 angles are 81$^\circ$ and 
100$^\circ$. 
Thus, it is best,when thinking about the RE site crystal field, to 
consider the true symmetry, -3m(D$_3$d), rather than relying on cubic or
axial approximations.

The known situation with respect to the presence or absence of long
range order in the RE$_2$Ti$_2$O$_7$ pyrochlores is summarized in 
Table 1.

One obvious correlation is that those RE$_2$Ti$_2$O$_7$ pyrochlores 
which
contain a Kramers (odd electron) ion nearly always show long
range order (Dy being the			 
exception) while those with a non-Kramers (even electron) ion do not.      
A ``zeroth order'' interpretation of the trends in Table 1. is then, that the
action of the relatively low symmetry crystal field induces a true singlet
ground state in the non-Kramers ions and this is the explanation of the 
absence of  long
range order (LRO). There is good evidence that such is the case for RE=Tm
from a combination of susceptibility~\cite{tm2ti2o7_1,tm2ti2o7_2},
inelastic neutron scattering~\cite{tm2ti2o7_2}, and
crystal field calculations (using the correct -3m symmetry)~\cite{faucher}. 
Experimentally~\cite{tm2ti2o7_2},
the singlet state is well-separated by 120K from the nearest excited state
which is in remarkable agreement with the aforementioned crystal field
calculations which predict 118K~\cite{faucher}.

The other two non-Kramers ions are not so simple. For 
Ho$_2$Ti$_2$O$_7$
the ground state is thought to be an Ising doublet~\cite{spin-ice},
in agreement with crystal-field calculations~\cite{faucher}, and the nearest
neighbor exchange is weakly ferromagnetic. Here it has been argued that
the strong Ising-like single ion anisotropy along the $[111]$
frustrates the development of long range
ferromagnetic order~\cite{ho2ti2o7,spin-ice}.
However, recent studies suggest a more complex 
picture where dipolar interactions competing with {\it antiferromagnetic} 
exchange is responsible for the behavior observed in Ho$_2$Ti$_2$O$_7$~\cite{shastry}. 
This material also exhibits spin dynamics and spin
freezing reminiscent of the disorder-free,intrinsic glassy behavior 
exhibited by the ``ice model''~\cite{ho2ti2o7,spin-ice}
with an exponential decrease of the spin
lattice relaxation rate suggestive of Orbach processes~\cite{orbach}.
In contrast  Gd$_2$Ti$_2$O$_7$ exhibits no apparent dynamics or 
spin-glassiness at any temperature even above $T_c$.

A detailed study of Tb$_2$Ti$_2$O$_7$ will be described in a subsequent
publication~\cite{tb2ti2o7_prop}. The salient facts are that the Tb$^{3+}$ 
ground state also
appears to be
a doublet but not so well isolated from several other levels within
15K$-$100K. The exchange interactions are relatively strongly
antiferromagnetic, comparable to Gd$_2$Ti$_2$O$_7$, and short range 
magnetic correlations persist up to at least 30K, also similar to what is 
found in Gd$_2$Ti$_2$O$_7$. Tb$_2$Ti$_2$O$_7$
does not order down to 70 mK~\cite{tb2ti2o7}.  The lack of LRO in this 
system is difficult to understand~\cite{tb2ti2o7_prop}.  Indeed, as argued in
refs~\cite{bramwell-obdo,moessner_cef}  a nearest-neighbor
Heisenberg antiferromagnet with a $[111]$ easy axis is a trivial problem
with an effectively non-frustrated and unique (two-fold Ising-like globally 
degenerate)
classical ground state, and should
therefore show a phase transition
at nonzero temperature in the limit of sufficiently strong crystal-field
level splitting compared to the superexchange $J$.
			
From the above discussion
we can conclude that each RE$_2$Ti$_2$O$_7$ material presents
its own special set of circumstances where details of the finely tuned 
relative strength of crystal field parameters,  exchange and dipolar 
couplings play a crucial role, and a blanket explanation for the 
apparent systematics of Table 1 will not be found. It is worth noting
an interesting paradox. Gd$_2$Ti$_2$O$_7$ represents  the case for which
some of the perturbations which might be thought to aid in the selection
of an unique ground state, i.e., crystal fields and anisotropy, are
largely absent, yet it orders. On the other hand, Tb$_2$Ti$_2$O$_7$ and 
Ho$_2$Ti$_2$O$_7$ in which crystal fields and anisotropy are clearly 
important, do not order and it is likely that these perturbations in fact inhibit 
the occurrence of long range order by competing with important interactions 
{\it other}~\cite{shastry} than nearest-neighbor Heisenberg antiferromagnetic 
exchange~\cite{bramwell-obdo,spin-ice,moessner_cef}.  
Some other interesting recent results have been found in 
the Yb$_2$Ti$_2$O$_7$ and Dy$_2$Ti$_2$O$_7$
pyrochlores~\cite{HoDyYb2Ti2O7_recent}.

It is also useful to compare Gd$_2$Ti$_2$O$_7$
with other Heisenberg
pyrochlores such as 
Y$_2$Mo$_2$O$_7$~\cite{dunsiger,gingras_ymoo,gardner_y2mo2o7}
and Heisenberg spinels
such as ZnFe$_2$O$_4$~\cite{potzel}.
Y$_2$Mo$_2$O$_7$ is a well-known 
geometrically frustrated antiferromagnet spin glass
material with $\theta_{\rm CW}/T_f \ge 10$, T$_f$ being the spin freezing 
temperature
of 21K~\cite{gingras_ymoo}.
Here too, only a speculation can be offered for the differences as
follows. Because of the high level of degeneracy across the zone, the 
nearest-neighbor Heisenberg pyrochlore antiferromagnet is expected to be 
fragile
against a small, random disorder level, $x$, and will have a propensity to
develop a {\it disorder-driven} spin glass ground state the smaller the 
perturbations $\{H'\}$
beyond the nearest-neighbor exchange interaction 
is~\cite{villain,bramwell-obdo}.
We expect that the 
critical disorder level for the N\'eel to spin-glass transition, $x_c$,
will go to zero as $\{H'\}$ goes to zero~\cite{villain}.
For example, in Y$_2$Mo$_2$O$_7$, there
is preliminary evidence that the second neighbor exchange parameter, 
J$_2$,
is only a few percent of $J_1$~\cite{raju_ymoo}. In Gd$_2$Ti$_2$O$_7$
on the other hand the leading
corrections $\{H'\}$  are dipolar interactions, $D_{dd}$, and of order 20\% 
of $J_1$.
In other words $\{H'\}/J_1$ is not small in Gd$_2$Ti$_2$O$_7$
and the anisotropy of the
dipolar interactions will possibly introduce sizeable stabilizing anisotropy
gaps to the spin-wave excitations out of the selected long range ordered 
ground
state. Both the relative size of $D_{dd}/J_1$ and the ``spin-holding'' effect
of the anisotropy of dipolar interactions will result in a much increased
$x_c$ compared to more isotropic Heisenberg systems with small 
$\{H'\}$~\cite{villain}.
In summary in this picture weak disorder drives the spin glass
transition in Y$_2$Mo$_2$O$_7$ while the strong and anisotropic dipolar
interactions ``helps'' stabilize long range order in Gd$_2$Ti$_2$O$_7$. In 
this context
the existence of a very weakly dispersive line along $[111]$ restored by 
order-by-disorder, or perturbative
$J_2$ and $J_3$, would suggest that, as in the fcc 
antiferromagnet~\cite{FCC}, weak random disorder would rapidly drive 
Gd$_2$Ti$_2$O$_7$ into a spin glass state. 

Finally, it is interesting to compare the behavior of Gd$_2$Ti$_2$O$_7$ with 
the frustrated ZnFe$_2$O$_4$ antiferromagnet spinel where Fe$^{3+}$ is 
a $^6S_{5/2}$ closed shell ion for which single-ion anisotropy should be 
negligible as is the case for the $^8S_{7/2}$ Gd$^{3+}$ ion in 
Gd$_2$Ti$_2$O$_7$ . 
In the insulating normal Heisenberg spinel ZnFe$_2$O$_4$, where the
Fe$^{3+}$ magnetic moments occupy a lattice of corner-sharing tetrahedra,
muon spin relaxation and neutron studies have revealed that 
long range antiferromagnetic order (LRO) develops 
below $T_{\rm N}=10.5$K. However, already
at temperatures of about $T\approx 10T_{\rm N}$
a short range antiferromagnetic order (SRO) develops
which extends through $\approx$70\%
of the sample volume just above $T_{\rm N}$. Below $T_{\rm N}$
antiferromagnetic SRO and LRO coexist. At 4.2K still $\approx 20\%$
of the sample are short range ordered. The
regions exhibiting SRO are very small $\approx 30\AA$.
The physical origin of the SRO as well as partial glassy behavior in 
ZnFe$_2$O$_4$ remains an enigma. Hence, while it appears that 
Gd$_2$Ti$_2$O$_7$ displays conventional antiferromagnetic long range 
order and 
Y$_2$Mo$_2$O$_7$ shows full-blown spin-glass behavior, 
ZnFe$_2$O$_4$ exhibits a combination of both short and
long range antiferromagnetic order, as well as spin-glassy behavior.
The origin of the difference between the Gd and Fe based pyrochlore lattice 
antiferromagnets in terms of their coexistence of long range order and
spin-glassy behavior is not known.  Possibly different range of interactions, 
presence of strong dipolar anisotropy in Gd$_2$Ti$_2$O$_7$ compared to 
a much more overall isotropic spin-spin interaction in ZnFe$_2$O$_4$ may 
play some role. In light of this, it would  be interesting to study in further 
detail the magnetic properties of Gd$_2$Ti$_2$O$_7$ using muon spin 
relaxation and neutron scattering methods.


\section{Conclusion} \label{conclusion}

Evidence has been presented from a.c. and d.c. suseptibility and
specific heat measurements, that the frustrated Gd$_2$Ti$_2$O$_7$
insulating pyrochlore 
exhibits a transition to a long range ordered state at 0.97K as opposed
to a spin glass or spin liquid state as often observed in other 
pyrochlore materials. From specific heat measurements, short range 
magnetic correlations have been found 
to extend to $T >30 T_c$ and the entropy removal
below $T_c$ is only about 50\%. From a mean field theoretical study it is
concluded that no long range order should exist for the pyrochlore lattice 
for nearest neighbor antiferromagnetic interactions, $J_1$, only,  even upon
inclusion of long range, anisotropic dipolar couplings, $D_{dd}$. Long 
range
order at  various commensurate or incommensurate wave vectors
is predicted to occur only upon including a finite second, $J_2$, and/or 
third, $J_3$, 
nearest-neighbor exchange interactions beyond $J_1$ and $D_{dd}$.
Long range order could also be driven by thermal and/or quantum 
fluctuations via
an order-by-disorder mechanism. The exact wave
vector depends on the relative signs and magnitudes of $J_2$, $J_3$ and
$D_{dd}$. It would be of interest to investigate further the nature of the
ordered state in zero and applied fields in Gd$_2$Ti$_2$O$_7$
by neutron scattering and muon spin relaxation methods. 
Finally, we argued above that the related fcc antiferromagnet material, 
cubic-Gd$_2$O$_3$, should be reconsidered as a geometrically frustrated 
antiferromagnet and is worthy of further study.


\section{Acknowledgements}

This work was supported by
NSERC operating grants, and also via an NSERC Collaborative Grant
on {\it Frustrated Antiferromagnets}.   M.G.  acknowledges Research
Corporation for a Research Innovation Award on ``Effect of Random
Disorder in Highly-Frustrated Antiferromagnets. We  thank S. Bramwell, B. 
den Hertog, M. Faucher, M. Harris, P.  Holdsworth, C. Lacroix, 
O. Petrenko, A. Ramirez and, specially,  J. Reimers for many useful and 
stimulating discussions.



\newpage

\begin{figure} 
\begin{center}
{\bf TABLE I}\\
\vspace{3mm}
{Presence or Absence of Long Range Order in RE$_2$Ti$_2$O$_7$
Pyrochlores.}

\vspace{3mm}

\begin{tabular}{||c|c|c|c||}    \hline\hline
\multicolumn{1}{||c}{\bf Rare Earth} &
\multicolumn{1}{|c}{$n$ in 4f$^n$} &
\multicolumn{1}{|c|}{\bf Long Range Order ($T_c$)} &
\multicolumn{1}{c||}{\bf Ref.}\\  \hline
      Gd      &           7      &      YES,(0.97K) & This work      \\
      Tb      &           8      &      NO             & Ref.~\cite{tb2ti2o7,tb2ti2o7_prop}\\
      Dy      &           9      &      NO             & 
Ref.~\cite{DyYbEr2Ti2O7,HoDyYb2Ti2O7_recent} \\
      Ho      &          10      &      NO             & 
Ref.~\cite{ho2ti2o7,HoDyYb2Ti2O7_recent} \\
      Er      &          11      &      YES,(1.25K)   & 
Ref.~\cite{DyYbEr2Ti2O7}\\
      Tm      &          12      &      NO             &  
Ref.~\cite{tm2ti2o7_1,tm2ti2o7_2}\\
      Yb      &          13      &               YES,(0.21K)    &
Ref.~\cite{DyYbEr2Ti2O7,HoDyYb2Ti2O7_recent}
\\ \hline\hline
\end{tabular}
\end{center}
\end{figure}

\vspace{2cm}

\begin{center}
{\bf  Figure Captions} \end{center}

\vspace{-1cm}
\begin{figure}
\begin{center}
  {
  \begin{turn}{0}%
    {\epsfig{file=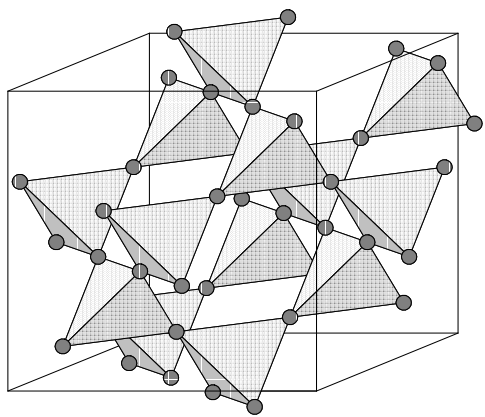,height=11.2cm,width=8cm} }
   \end{turn}
   }
\vspace{-4cm}
\caption{Pyrochlore lattice of corner-sharing tetrahedra.}
\end{center}
\end{figure}

\newpage

\vspace{2cm}

\begin{figure}
\begin{center}
  {
  \begin{turn}{0}%
    {\epsfig{file=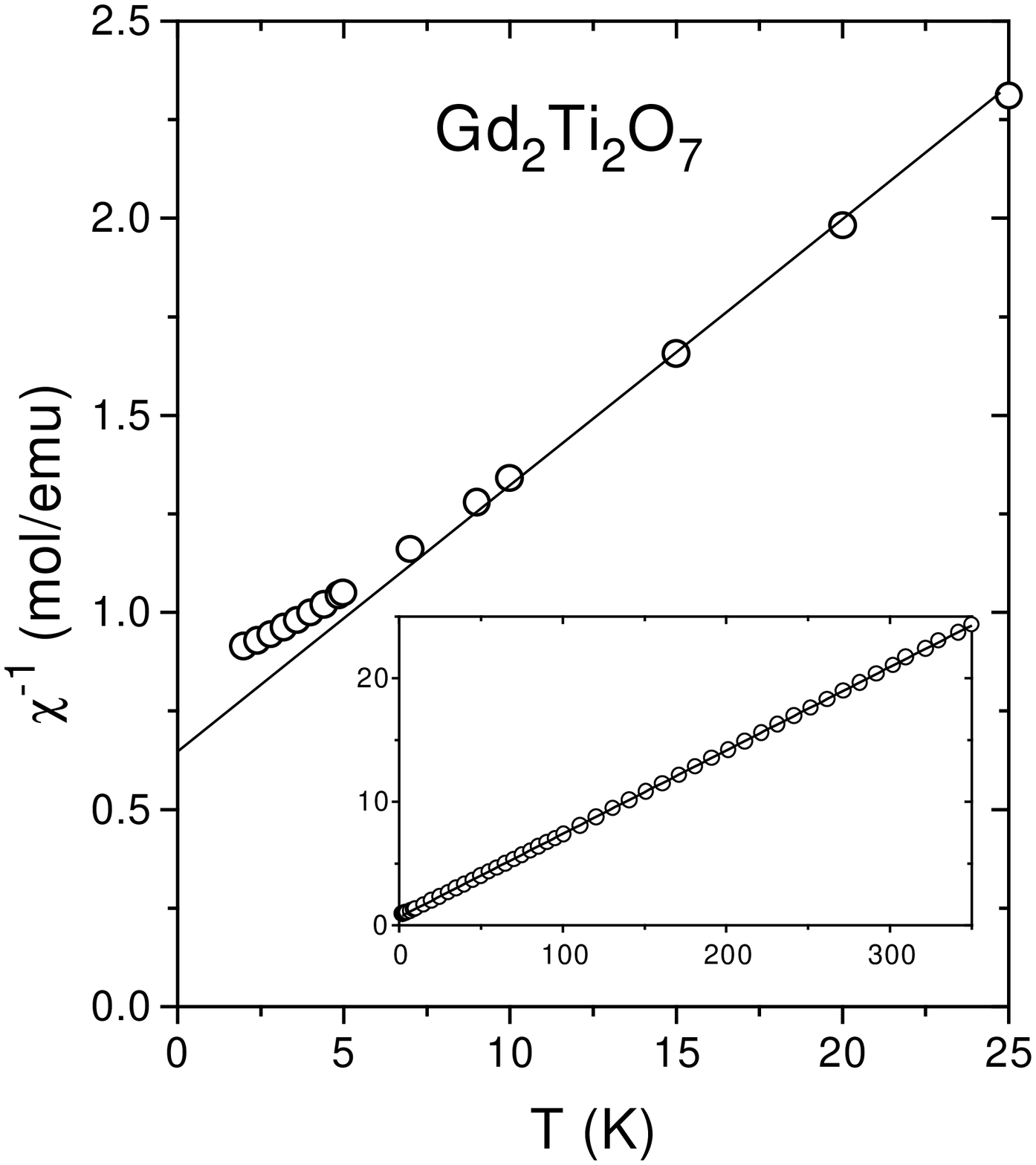,height=8cm,width=8cm} }
   \end{turn}
   }
\vspace{1cm}
{
  \begin{turn}{0}%
    {\epsfig{file=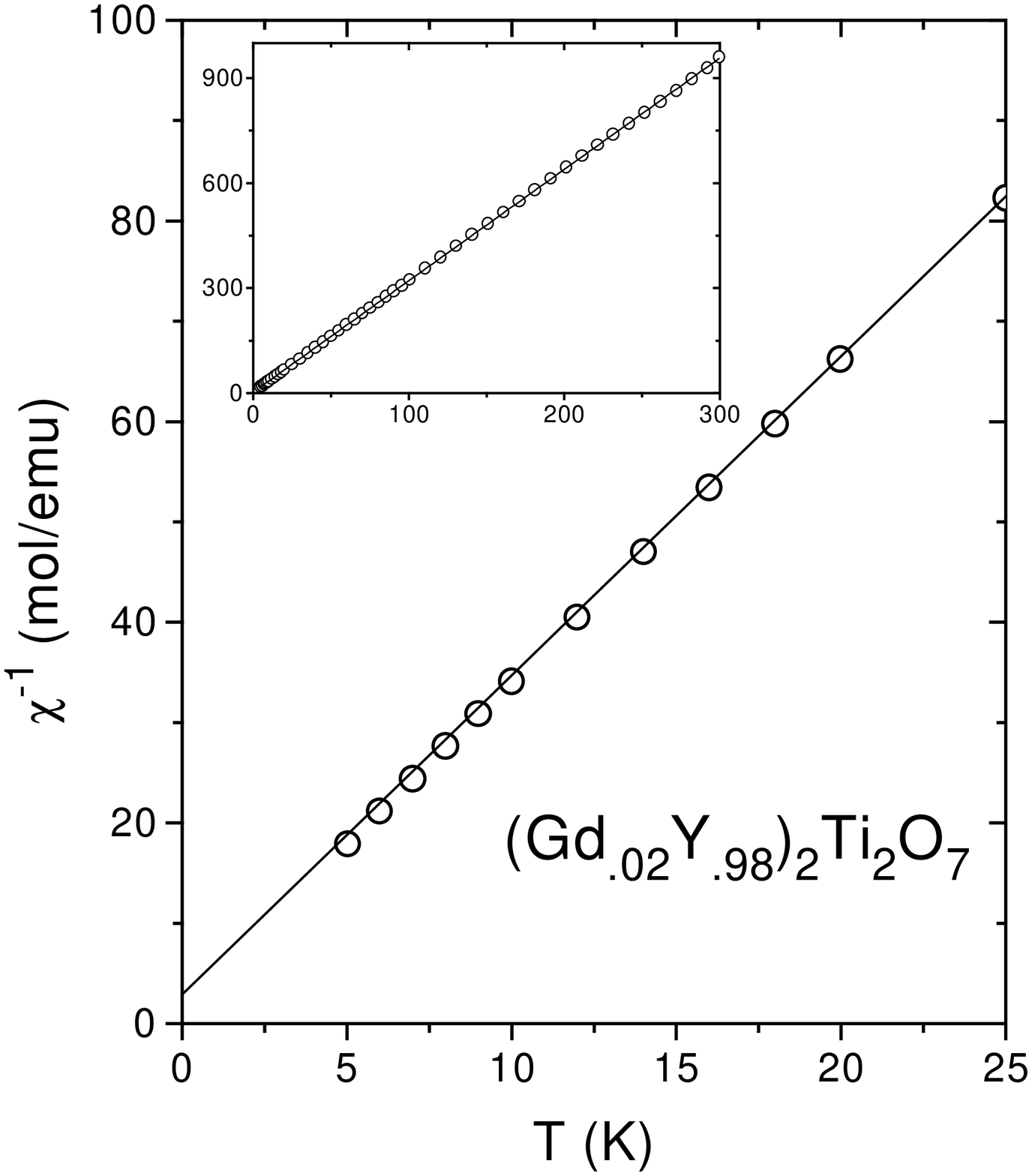,height=8cm,width=8cm} }
   \end{turn}
   }
\caption{(a) Inverse molar susceptibility, $1/\chi$,  of
Gd$_2$Ti$_2$O$_7$ against temperature in the temperature range $T=[2-
25]$ K, and in the temperature range $T=[2-300]$ K in the inset.
(b) Inverse
molar susceptibility, $1/\chi$, of
(Gd$_{0.02}$Y$_{0.98}$)$_2$Ti$_2$O$_7$ against temperature in the
temperature range $T=[2-25]$ K, and in the temperature range $T=[2-300]$
K in the inset.}
\end{center}
\end{figure}


\begin{figure}
\begin{center}
  {
  \begin{turn}{0}%
    {\epsfig{file=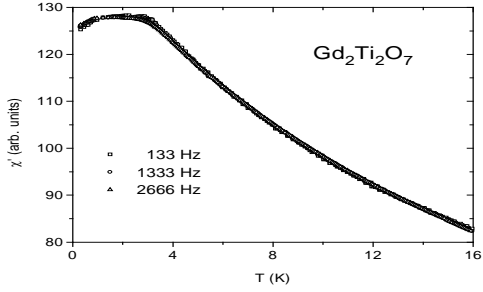,height=8cm,width=8cm} }
   \end{turn}
   }
\caption{Real
part of the ac susceptibility, $\chi'$, vs. temperature measured
at different frequencies.}
\end{center}
\end{figure}


\begin{figure}
\begin{center}
  {
  \begin{turn}{0}%
    {\epsfig{file=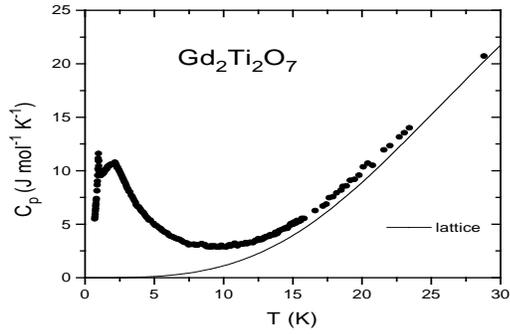,height=8cm,width=8cm} }
   \end{turn}
   }
\vspace{-1cm}
\caption{Specific heat, $C_p$, of  Y$_2$Ti$_2$O$_7$
as function of temperature. The solid
line corresponds to the lattice specific heat, $C_l$, estimated from
the measurements on the non-magnetic Y$_2$Ti$_2$O$_7$.}
\end{center}
\end{figure}

\newpage


\begin{figure}
\begin{center}
\vspace{1cm}
\caption{Magnetic specific heat, $C_m$,  
(obtained by subtracting $C_l$ from
$C_p$)
against temperature. The solid line represents the theoretical fit (see
text for details). The inset shows a blown-up region of $C_m$ in the 
low-temperature regime.}
\end{center}
\end{figure}

\vspace{2cm}

\newpage


\begin{figure}
\begin{center}
  {
  \begin{turn}{0}%
    {\epsfig{file=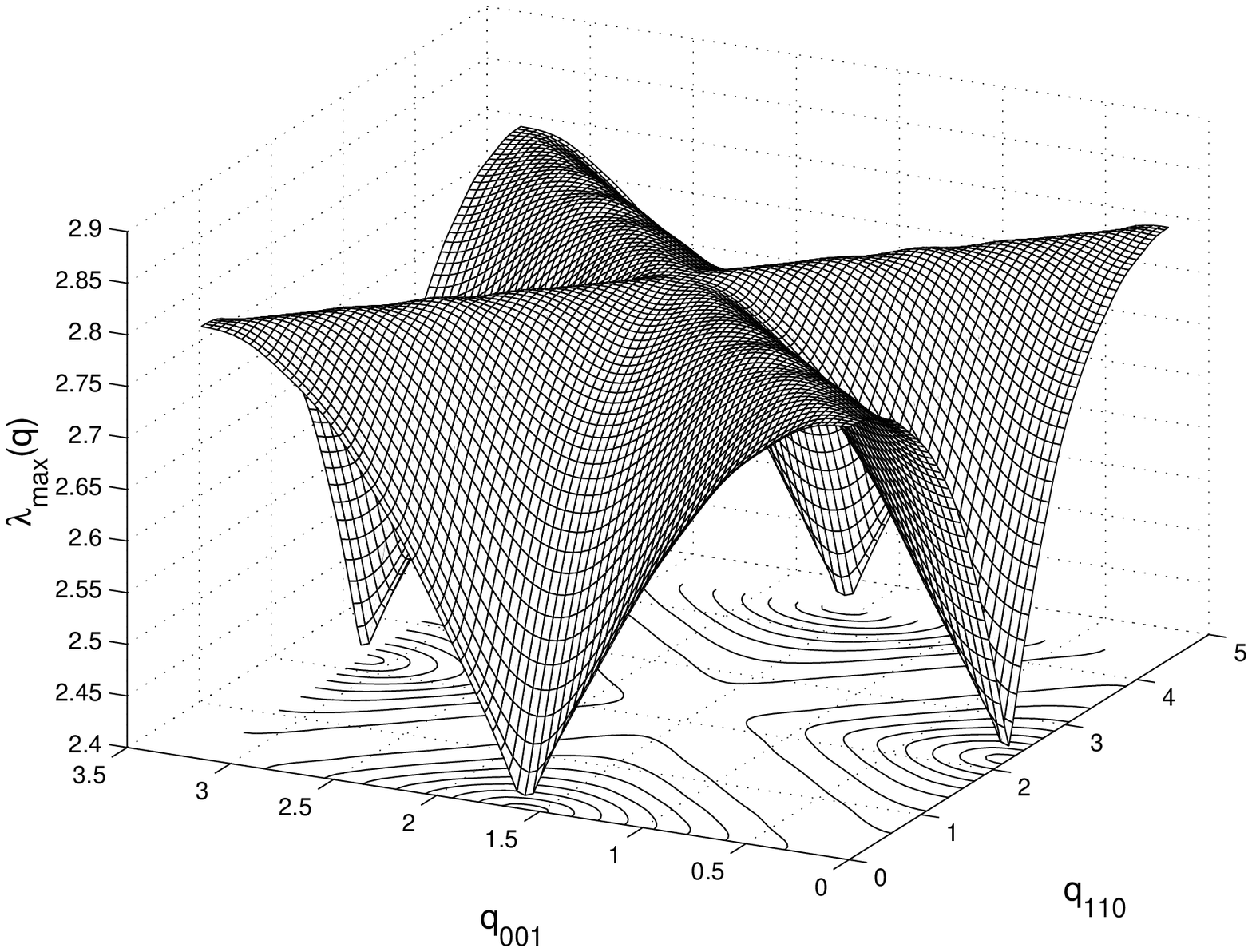,height=14cm,width=14cm} }
   \end{turn}
   }
\caption{Largest eigenvalue $\lambda_{\rm max}({\bf
q})$ as a function
of wavevector ${\bf q}$ for $D_{dd}/J_1=0.2$ and $J_2=J_3=0$.
A degeneration line occurs for ${\bf q}$ in the star of the $[111]$ direction.
The small ``ripples'' seen on the degeneration lines along $(111)$ and
$(11\bar 1)$ directions are due to the finite number (500) nearest-neighbors
considered in the dipolar interactions. When considering more than 10
nearest-neighbors,
the maximum of  $\lambda({\bf q})$ always occurs on the star of $[111]$
with
the amplitude of the modulations due to the dipolar cut-off continuously
decreasing as
the number of nearest-neighbors is increased to infinity.}
\end{center}
\end{figure}


\begin{thebibliography}{99}
\bibitem{dipole_theta_cw}To obtain a more reliable estimate of
$D_{dd}/J_1$, one should determine the Curie-Weiss contribution
from dipole-dipole interactions to the high-temperature limit of
$\chi$, and hence $\theta_{\rm CW}$,
in order to extract a more precise estimate of $J_1$. 
The results of such calculations will be produced 
elsewhere~\cite{tb2ti2o7_prop}.
Note that the specific value chosen for
$D_{dd}/J_1$ with $J_2=J_3=0$ does not affect the
results of a line of degeneracy along the star of $[111]$
(Ref.~\cite{dion_gd2ti2o7}).
\bibitem{range_dipoles}We have found that the results for the selected 
wavevector(s) at the mean-field level depends sensitively on the range of 
dipolar interactions considered. Here, we have included the first 500 
nearest-neighbors in the dipolar sum.
\end{thebibliography}
\end{document}